\documentclass[aps,pre,12pt,notitlepage,superscriptaddress,nofootinbib]{revtex4-2}

\usepackage{amsmath,graphicx}
\usepackage{tabularray}
\usepackage{hyperref,times}
\hypersetup{
    colorlinks=true,       
    linkcolor=blue,          
    citecolor=blue,        
    filecolor=blue,      
    urlcolor=blue          
}
\usepackage{upgreek}

\newcommand{\vecA}{{\cal A}}
\newcommand{\vecD}{{\cal D}}
\newcommand{\vecM}{{\cal M}}
\newcommand{\vecX}{{\bf X}}
\newcommand{\vecf}{{\bf f}}
\newcommand{\vecn}{{\hat{\bf n}}}
\newcommand{\vecr}{{\bf r}}
\newcommand{\vecq}{{\bf q}}
\newcommand{\vecv}{{\bf v}}
\newcommand{\eg}{{e.g., }}
\newcommand{\ie}{{i.e., }}
\newcommand{\tr}{{\rm tr}}

\usepackage{xcolor}

\begin{document}

\title{Detecting and characterizing phase transitions in active matter using entropy}

\author{Benjamin Sorkin}

\affiliation{School of Chemistry and Center for Physics and Chemistry of Living Systems, Tel Aviv
  University, 69978 Tel Aviv, Israel}

\author{Avraham Be'er}

\affiliation{Zuckerberg Institute for Water Research, The Jacob Blaustein Institutes for Desert Research, Ben-Gurion University of the Negev, Sede Boqer Campus 84990, Midreshet Ben-Gurion, Israel.}
\affiliation{
Department of Physics, Ben-Gurion University of the Negev, 84105 Beer Sheva, Israel.
}

\author{Haim Diamant}

\affiliation{School of Chemistry and Center for Physics and Chemistry of Living Systems, Tel Aviv
  University, 69978 Tel Aviv, Israel}

\author{Gil Ariel}
\email{arielg@math.biu.ac.il}

\affiliation{Department of Mathematics, Bar-Ilan University, 52000
  Ramat Gan, Israel}

\begin{abstract}

A major challenge in the study of active matter lies in quantitative characterization of phases and transitions between them. We show how the entropy of a collection of active objects can be used to classify regimes and spatial patterns in their collective behavior. Specifically, we estimate the contributions to the total entropy from correlations between the degrees of freedom of position and orientation. This analysis pin-points the flocking transition in the Vicsek model while clarifying the physical mechanism behind the transition. When applied to experiments on swarming {\it Bacillus subtilis} with different cell aspect ratios and overall bacterial area fractions, the entropy analysis reveals a rich phase diagram with transitions between  qualitatively different swarm statistics. We discuss physical and biological implications of these findings.

\end{abstract}

\maketitle

\section{Introduction}
\label{sec:intro}

Active matter is composed of motile objects which continuously consume energy to generate propulsion~\cite{Rev:MarRam12,Rev:Ram10}. Such systems have been extensively studied in the past two decades.  They are of particular relevance to biology, describing the collective motions of a wide variety of entities~\cite{vicsek2012collective}, ranging from swarming bacteria~\cite{GilBacteria20,be2019statistical} and migrating cells~\cite{mehes14bacteria} to flocks of birds~\cite{ballerini08birds}, marching locusts~\cite{buhl06locusts}, and human crowds~\cite{helbing2001self}. Numerous types of inanimate particles, such as light-activated colloids~\cite{palacci2013photocolloids} and chemically-activated diffusiophoretic particles~\cite{robertson2018diffusiophorhetic}, exhibit similar behaviors.

With its perpetual energy dissipation, active matter provides examples of far-from-equilibrium steady states with intriguing statistical-mechanical aspects~\cite{Rev:MarRam12,Rev:Ram10,Rev:Ram17,fodor2016EP}. These include: (a) breakdown and modification of the fluctuation-dissipation theorem~\cite{kikuchi09FDT,fodor2016EP,caprini2021FDT}; (b) giant number fluctuations and breakdown of the Mermin-Wagner theorem~\cite{Rev:Ram10,Rev:MarRam12,zhang2010GNF}; and (c) the inadequacy of  temperature~\cite{loi2008temp,takatori2015thermo}, pressure~\cite{solon2015pressure,takatori2014pressure}, and chemical potential~\cite{takatori2015thermo,Rev:Ram17} as thermodynamic state variables. 

In contrast, entropy remains well-defined out of equilibrium through its connection to information. The entropy quantifies the information content of the system's statistics through Shannon's formula~\cite{Shannon1948},
\begin{equation} 
  H=-\sum_{\vecX}p_{\vecX}\ln p_{\vecX}=-\int d\vecX\, p(\vecX)\ln p(\vecX),
\label{eq:shannon}
\end{equation}
where $p_{\vecX}$ ($p\left(\vecX\right)$) is the probability (density function) to obtain the discrete (continuous) microstate $\vecX$.
Entropy-based considerations have been  effective for identifying and characterizing changes in material properties, e.g., in equilibrium phase transitions~\cite{kardarfields,frenkel99rev,deGennesBook}, pattern formation~\cite{PatternFormRev}, and self-assembly~\cite{SelfAsRev}. Yet, they have rarely been applied to active systems. The main reason is seemingly technical\,---\,estimation of entropy out of equilibrium has proven challenging. 
Direct sampling of $p(\vecX)$ to be used in Eq.~(\ref{eq:shannon}) is often either inapplicable or unreliable in high-dimensional, continuous phase spaces (see, for example, Refs.~\cite{Ariel2020,paninski2003,beirlant1997}). 
As a result, there is an ongoing effort to find reliable entropy estimators, in particular, for experimental data~\cite{Zu2020,AvineryPRL2019,MartinianiPRX2019,NirPNAS2020,ArielPRE2020,SorkinPRE22}. 

Several earlier works considered aspects of entropy (or effective free energy) of active systems. Some involved coarse-grained theories based on stochastic equations~\cite{barre2015cont_ent,stenhammar2013cont_ent,tailleur2008cont_ent,speck2014cont_ent}. Another work used entropy estimation of interlaced frames using the lossless-compression method~\cite{cavagna2021ent}, thus also incorporating temporal information. Lastly, a dynamical inference approach addressed both the static and dynamic (caliber) entropies, which were maximized under the constraint of a given orientations correlation function (CF)~\cite{cavagna2014ent}. Interestingly, in the latter, the estimation based only on steady-state entropy (relying on spatial correlations) failed to describe real bird flocks.
The current work builds on the structure-based entropy inference presented in Refs.~\cite{ArielPRE2020,SorkinPRE22}. The method has been successfully applied to several non-equilibrium systems. Here we apply it to two active systems.

Recent works on active matter have dealt with entropy \emph{production} (\eg Refs.~\cite{fodor2016EP,nardini2017EP,SpeckEL2016,shankar2018EP,guo2021EP}). While entropy is a state property, not requiring a unit of time, entropy production is a dynamic quantity related to the heat dissipation per unit time. Out of equilibrium energy may constantly flow into the system, and heat dissipation (\ie entropy production) is necessary to sustain a steady state. At that steady state, properties such as the entropy are constant. 
In this context, other forms of dynamical entropy approaches are worth mentioning, such as estimating the $\epsilon$-entropy and the Kolmogorov-Sinai entorpy~\cite{AbelPD00}, the maximum-caliber principle~\cite{Jaynes80,cavagna2014ent}, and using the entropy as a Lyapunov function~\cite{Tadmor2008}.
Here we concentrate only on the steady-state entropy, associated with the steady-state distribution of microstates (\eg positions and orientations).

Significant theoretical and experimental efforts have been devoted to obtaining phase diagrams of active matter, \ie characterizing the observed regimes and transitions between them.
In most cases an appropriate order parameter (OP) is defined and measured or calculated. However, especially in active matter, a single OP or a few OPs may not capture the entire behavior of interest. For example, common OPs relate to the particles' orientations or velocities and do not reflect the details of their spatial arrangement, let alone more subtle cross-correlations between orientations and positions. This is despite the fact that spatial heterogeneities are known to play a major role in active systems~
\cite{Rev:Ram10,Rev:MarRam12,Rev:Ram17}. 
Indeed, different observables and statistical attributes, such as cluster size distributions, have been used. Such attributes (\eg the cluster size) are not always sharply defined~\cite{huepe2004clust,peruani2010clust,zhang2010GNF,GilBacteria20}.

Correlation functions are well-defined higher-level descriptors of particle arrangement compared to mean OPs. They are applicable out of equilibrium and routinely extracted from experiments and simulations.
We focus on the  information content (entropy) of CFs, following the approach of Refs.~\cite{SorkinPRE22,ArielPRE2020}. The idea is to measure or compute different pair correlation functions  corresponding to different degrees of freedom (DOFs) of the particles. Using these (Fourier-transformed) CFs as  constraints, we apply the functional developed in Refs.~\cite{SorkinPRE22,ArielPRE2020} to obtain an upper bound for the system's entropy. In other words, we estimate how much information has been gained by measuring the CF.
By constraining different CFs, or their combinations, one can identify the DOFs that dominate a certain regime or transition~\cite{SorkinPRE22}.
We will specifically use the two-point position CF (\ie the structure factor) and the orientations CF.

In the following Sec.~\ref{sec:methods} we 
specialize the method to the systems studied here\,---\,particles moving in two dimensions (2D) whose DOFs are position and orientation (either polar or nematic). 
In Sec.~\ref{sec:vicsek} we apply the method to the Vicsek model~\cite{vicsek1995model}, a canonical minimal model where the particles are cognizant of their neighbors' velocity direction in the presence of uniform noise. This model is known to exhibit a flocking transition with decreasing noise.
In Sec.~\ref{sec:bacteria} we use the method to re-analyze experiments on swarming bacteria~\cite{GilBacteria20}. The experiments have shown that elongated mutants undergo a swarming transition with qualitatively different dynamics compared to the native, less elongated bacteria. We discuss our findings and future perspectives in Sec.~\ref{sec:discuss}.
  
\section{Method}
\label{sec:methods}

The two systems of interest consist of $N$ particles in 2D, whose DOFs are their positions $\{\vecr_n=(x_n,y_n)\}$ and orientations $\{\theta_n\}$. Both systems are known to exhibit transitions involving changes in the spatial and orientational arrangements of the particles. 

For the positions, the typical two-point CF is the structure factor,
\begin{equation}
    S (\vecq)=\left\langle\frac{1}{N}\sum_{n=1}^N\sum_{m=1}^Ne^{-i\vecq\cdot(\vecr_n-\vecr_m)}\right\rangle,\label{eq:S(q)}
\end{equation}
which is the Fourier-transformed two-point correlation of the particle number density.
The angular brackets denote an ensemble average (over samples), and the factor $N^{-1}$ ensures that $S(\vecq)=1$ in the ideal-gas limit (uncorrelated positions). This factor appears inside the brackets  to treat cases where the number of particles varies between samples (as in Sec.~\ref{sec:bacteria}).
The wavevector $\vecq$ is discretized by the system's size (box of side $L$) according to $\sum_\vecq(\cdot)=(L/2\pi)^2\int d\vecq(\cdot)$. The uniform mode ($\vecq=\mathbf{0}$) is excluded throughout. 

Similar to the case of particle number density, one can define a two-point correlation of the orientation density field. The angle DOF can be transformed using spherical harmonics, which in 2D are $Y_l(\theta)=e^{il\theta}$, $l=0,1,\dots$. In this sense, $S (\vecq)$ is the $0$th harmonic correlator. The next harmonic is $l=1$ for a polar DOF ($\theta\in\left[-\pi,\pi\right)$), and $l=2$ for a nematic DOF ($\theta\in\left[-\pi/2,\pi/2\right)$). The orientations CFs are then $2\times2$ matrices,
\begin{equation}
     \vecD^l (\vecq) = \left\langle\frac{2}{N}\sum_{n=1}^N\sum_{m=1}^Ne^{-i\vecq\cdot(\vecr_n-\vecr_m)}\vecn (l\theta_n)\vecn^\dag (l\theta_m)\right\rangle,\label{eq:D(q)}
\end{equation}
where $\vecn(\theta)=(\cos\theta,\sin\theta)$ and $\dag$ denotes transpose and complex conjugation. The prefactor $2N^{-1}$ ensures that $\vecD^l (\vecq)=\mathcal{I}$ for an ideal gas ($\mathcal{I}$ being the identity matrix).

Our goal is to find the information content encoded in the CF. This is the minimal excess Shannon entropy with respect to the ideal gas (\ie uniform distribution with the same mean density) that any system can have given the measured CF. In other words, we find the maximum-entropy model with the CF as a constraint.
Following Ref.~\cite{SorkinPRE22}, we denote the excess entropy per particle, $h_\mathrm{ex}=\langle(H_N-H^\mathrm{id}_N)/N\rangle$, where $H_N$ is the system's entropy for $N$ particles, $H_N^\mathrm{id}$ is the ideal gas entropy for $N$ particles, and the average $\langle\cdot\rangle$ here is over the number of particles (for example, in analyzing the experiments with swarming bacteria below or in grand-canonical ensembles).

For homogeneous systems (\ie no symmetry-breaking external field), the upper bound of $h_\mathrm{ex}$ using the structure factor as the CF is given by \cite{ArielPRE2020}
\begin{equation}
    h_\mathrm{ex}[S]=\frac{1}{2\langle N\rangle} \sum_{\vecq\ne\mathbf{0}} \left[ \ln S (\vecq) + 1 - S(\vecq) \right].
    \label{eq:hex_S}
\end{equation}
Similarly, the entropy's upper bound as a functional of the orientations CF is \cite{SorkinPRE22}
\begin{equation}
    h_\mathrm{ex}[\vecD^l]=\frac{1}{2\langle N\rangle} \sum_{\vecq\ne\mathbf{0}} \left[ \ln\det\vecD^l (\vecq) + 2 - \tr\vecD^l (\vecq) \right].
    \label{eq:hex_D}
\end{equation}
The lower of the two entropy bounds given by Eqs.~(\ref{eq:hex_S}) and (\ref{eq:hex_D}) should point to the dominant DOF.

We will also use a mixed CF defined as the $3\times3$ matrix,
\begin{equation}
    \vecM (\vecq)=\vecA^{-1}\left\langle\frac{1}{N}\sum_{n=1}^N\sum_{m=1}^Ne^{-i\vecq\cdot(\vecr_n-\vecr_m)}\vecf(l\theta_n) \vecf^\dagger(l\theta_m)\right\rangle,\label{eq:hex_M}
\end{equation}
where $\vecf(\theta)=(1,\cos\theta,\sin\theta)$, and $\vecA^{-1}=\mathrm{diag}(1,2,2)$ is the normalization matrix ($\vecA$ being the ideal-gas limit of the expression in angular brackets). The corresponding upper bound for the entropy is
\begin{equation}
    h_\mathrm{ex}[\vecM]=\frac{1}{2\langle N\rangle} \sum_{\vecq\ne\mathbf{0}} \left[ \ln \det\vecM (\vecq) + 3 - \tr\vecM (\vecq) \right].
\end{equation}
Note that $M_{11}(\vecq)$ is the structure factor $S (\vecq)$, while the block $M_{ij}$, $i,j=2,3$, 
forms $\vecD^l (\vecq)$. The off-diagonal terms, $M_{12}(\vecq)=M_{21}^*(\vecq)$ and $M_{13}(\vecq)=M_{31}^*(\vecq)$, are the position-orientation cross-correlations. 
Since the spherical harmonics form an orthogonal basis, $h_\mathrm{ex}[\vecM]$ is guaranteed to be smaller than both $h_\mathrm{ex}[S]$ and $h_\mathrm{ex}[\vecD^l]$. The maximum $h_\mathrm{ex}[\vecM]$, obtained for uncorrelated positions and orientations, is $h_\mathrm{ex}[S]+h_\mathrm{ex}[\vecD^l]$. One may similarly include more CFs corresponding to additional spherical harmonics, which will obviously reduce the entropy bound further.

\section{Vicsek model}
\label{sec:vicsek}

\begin{figure}
\centering
\includegraphics[scale=0.33]{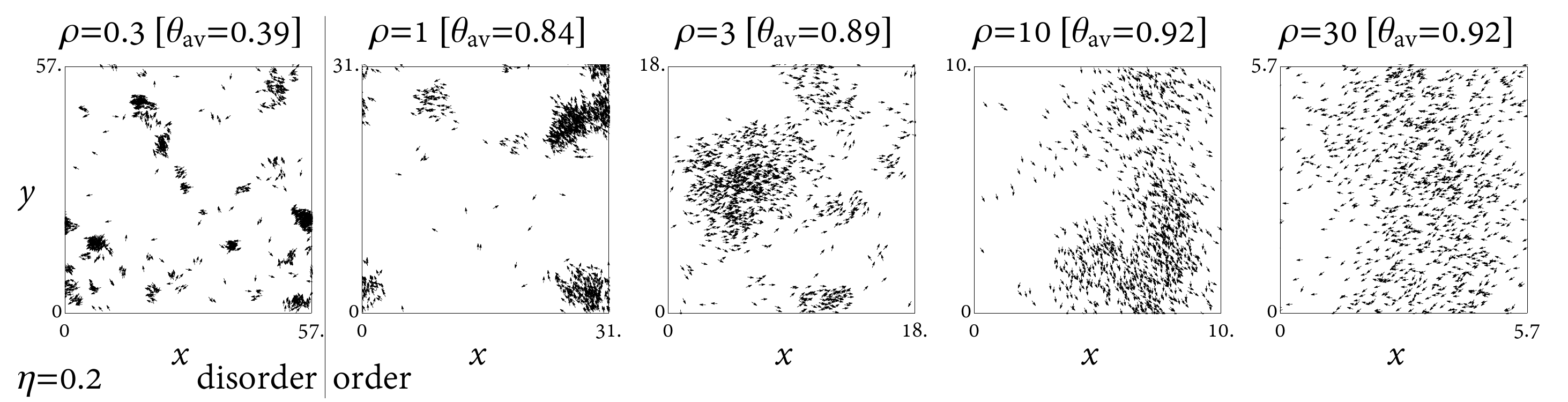}
\includegraphics[scale=0.33]{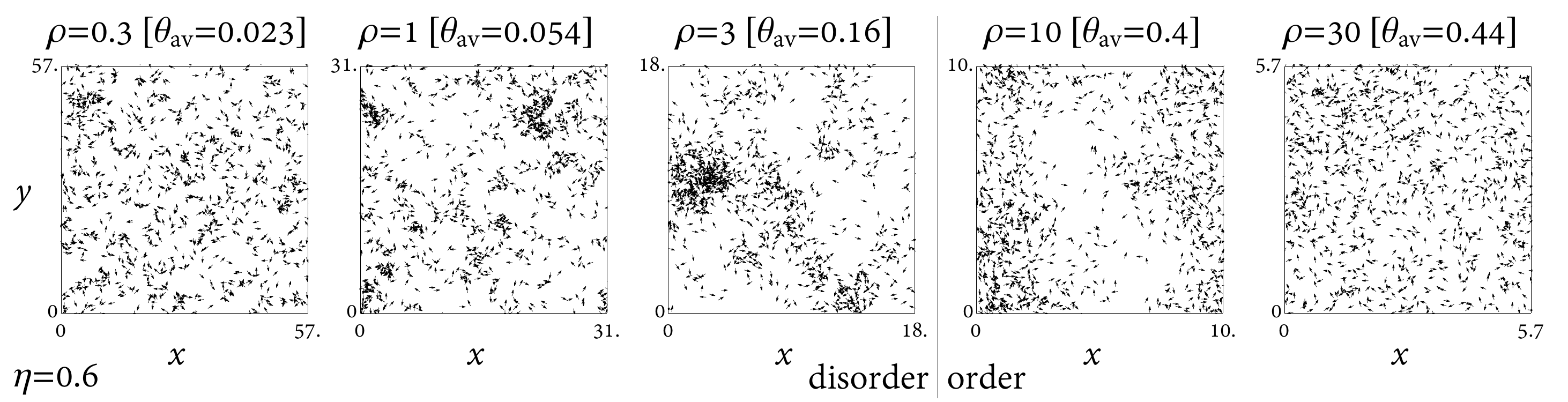}
\caption{Illustrative snapshots from simulations of the Vicsek model. The shown system contains $10^{3}$ particles under constant noise (top row: $\eta=0.2$, bottom row: $\eta=0.6$), interaction radius ($r=1$), and velocity ($v=0.1$), and varying density $\rho$ as indicated. $\theta_\mathrm{av}$ is the order parameter. With increasing density, the (spatial) flocking and (orientational) alignment become more pronounced. For lower noise the transition occurs earlier. See Fig.~\ref{fig:hex vs rho} for the corresponding entropy contributions. The vertical lines separate the orientationally disordered and ordered phases as we infer from Fig.~\ref{fig:hex vs rho}. Without the entropy bounds of Fig.~\ref{fig:hex vs rho}, the choice of where to put these `transitions lines' would have been uncertain.} 
\label{fig:Snaps}
\end{figure}

The Vicsek model \cite{vicsek1995model} is a minimal model for active matter, in which the dynamics of self-propelled objects leads to collective behavior and flocking. See Fig.~\ref{fig:Snaps} for typical snapshots. It is an example of the interplay between local alignment interactions and disordering noise. In this section we apply the method described in Sec.~\ref{sec:methods} to simulations of the Vicsek model. We obtain the excess entropy for different values of system parameters and resolve the role of each DOF in the collective dynamics. 

In the Vicsek model, particle velocities $\{\vecv_n\}$ have a fixed magnitude $v$, such that $\vecv_n=v(\cos\theta_n,\sin\theta_n)$. Initially (at $t=0$), particle positions are uniformly distributed on a square $L\times L$ with periodic boundary conditions. At each simulation time step, the orientation is updated to the direction of the average velocity of all neighbors up to an interaction distance $r$, $\tan[\overline{\theta_n(t)}]=\langle\sin(\theta_m(t))\rangle_{|\vecr_n-\vecr_m|<r}/\langle\cos(\theta_m(t))\rangle_{|\vecr_n-\vecr_m|<r}$, plus added (intrinsic) noise $\eta_n(t)$, uniformly distributed on $[-\eta\pi,\eta\pi]$, with $\eta\in[0,1]$. Positions are updated deterministically given the velocities. The equations of motion are summarized as
\begin{subequations}
\begin{eqnarray}
 \theta_n(t+1)&=&\overline{\theta_n(t)}+\eta_n(t),\\
 \vecr_n(t+1)&=&\vecr_n(t)+\vecv_n(\theta_n(t)).
\end{eqnarray}
\end{subequations}
The control parameters are the density, $\rho=N/L^2$, and the temperature-like noise $\eta$ (where $\eta=0$ corresponds to $T=0$, and $\eta=1$ to $T\to\infty$). Throughout this section, the values of $r=1$ and $v=0.1$ are fixed, and the density is controlled by changing $L$ at fixed $N$.

The order parameter (OP) is defined as $\theta_\mathrm{av}=|\langle e^{i\theta}\rangle|$, with $\theta_\mathrm{av}=1$ corresponding to perfect orientational order, and $\theta_\mathrm{av}=0$ to complete disorder. Upon varying $\rho$ and $\eta$, the OP shows a transition which sharpens with increasing $N$ and $L$ at constant $\rho$~\cite{vicsek1995model,vicsek07model}. Most transitions in active systems involve not only  alignment but also clustering~\cite{Rev:MarRam12,Rev:Ram10}. The OP as defined above, however, does not capture the spatial arrangement of particles. Common remedies are to define clusters and study their size distribution \cite{huepe2004clust,peruani2010clust}, or consider various CFs~\cite{ro2021disorder,pearce2021orientational,sanchez2012spontaneous}.

We performed simulations of the Vicsek model with $N=10^4$ particles and 100 runs for a range of densities and noise amplitudes. From each run we took 10 snapshots interspaced by more than the initial relaxation time, resulting in a total of $10^3$ samples. Each sample $s$ has the configuration $\{x_n^s,y_n^s,\theta_n^s\}$ ($n=1,\ldots,10^4$). From these samples we compute three CFs according to Eqs.~\eqref{eq:hex_S}-\eqref{eq:hex_M}: (a) the structure factor, $S(\vecq)$; (b) the polar orientations CF, $\vecD^1 (\vecq)$; and (c) the ``mixed" CF, $\vecM(\vecq)$, imposing simultaneously both $S(\vecq)$ and $\vecD^1(\vecq)$ and including also position-orientation cross-correlations.
Note that for $K$ samples with $N$ particles, all three CFs can be computed in ${\mathcal O} (NK)$ efficiency. In particular, the double sums over $N$ can be avoided by calculating for each sample $\sum_n e^{-i \vecq \cdot \vecr_n}$, $\sum_n e^{-i \vecq \cdot \vecr_n} \vecn(l \theta_n)$ or $\sum_n e^{-i \vecq \cdot \vecr_n} \vecf(\theta_n)$, and multiplying by the transposed conjugate.

\begin{figure}
\centering
\includegraphics[scale=0.77]{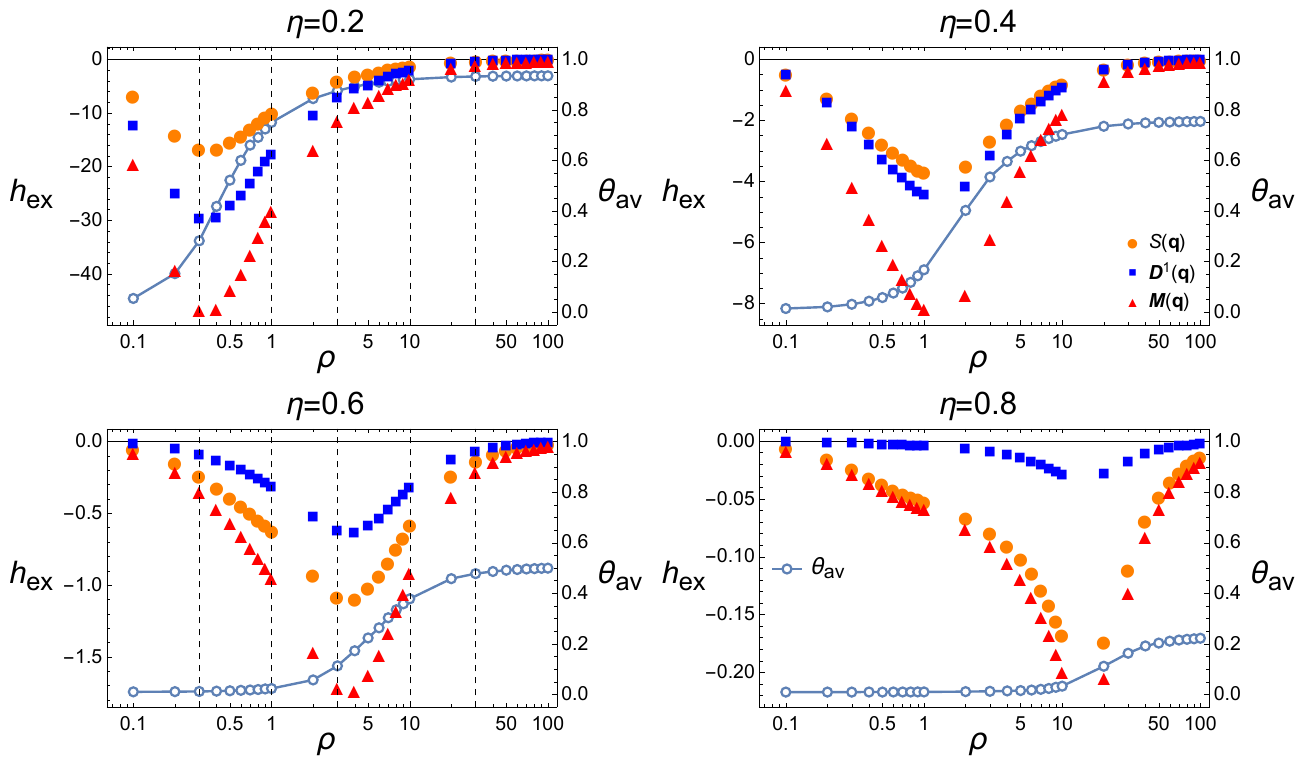}

\caption{\label{fig:hex vs rho}Excess entropy per particle $h_\mathrm{ex}$ in the Vicsek model as a function of density $\rho$ for  several noise amplitudes $\eta$. Different symbols show $h_\mathrm{ex}$ as obtained from different correlation functions: structure factor $S(\vecq)$ (orange circles), polar orientation correlation function $\vecD^1(\vecq)$ (blue squares), and mixed correlation function $\vecM(\vecq)$ (red triangles). The connected empty circles show the order parameter, $\theta_\mathrm{av}$. Dashed vertical lines in the left panels indicate the densities of the snapshots appearing in Fig.~\ref{fig:Snaps}. Parameters: $N=10^4$,  $r=1$, $v=0.1$.}
\end{figure}

\begin{figure}
\centering
\includegraphics[scale=0.77]{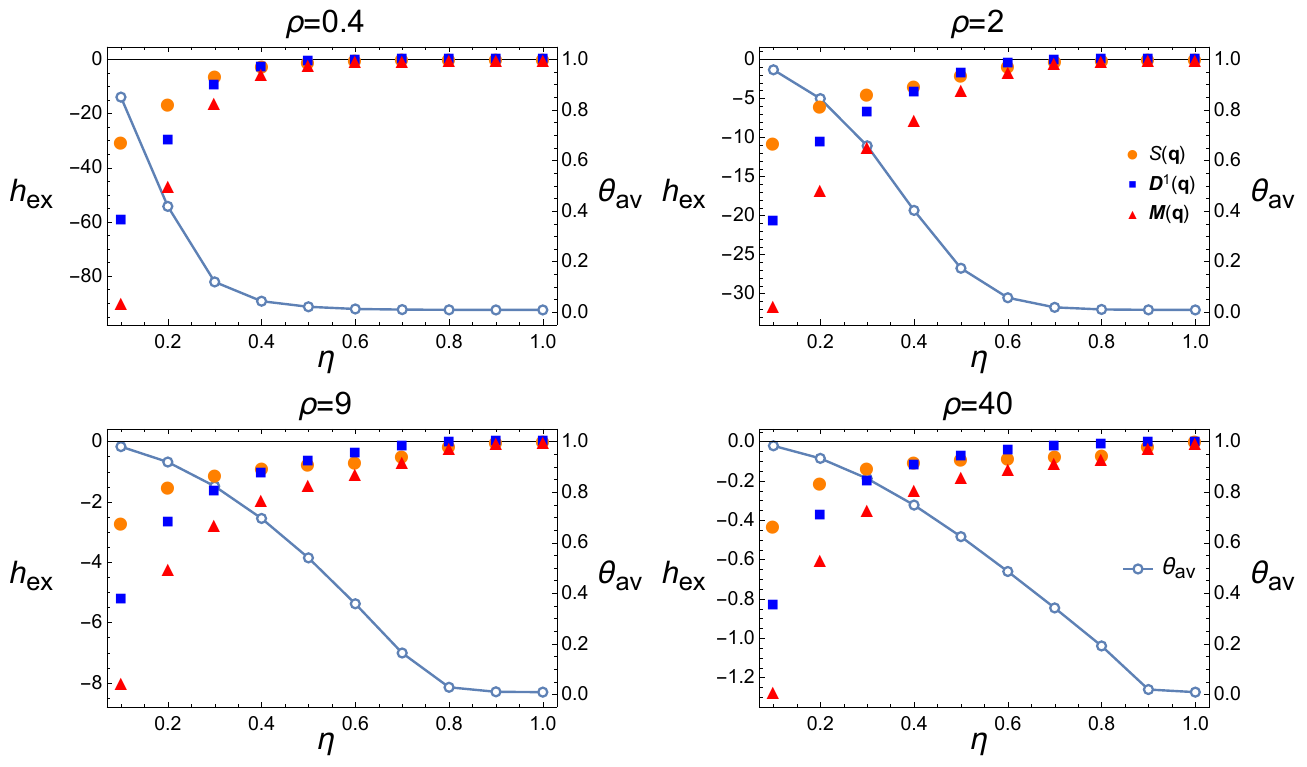}

\caption{\label{fig:hex vs eta}Excess entropy per particle in the Vicsek model as a function of noise amplitude for several densities. Symbols and parameters are the same as in Fig.~\ref{fig:hex vs rho}.}
\end{figure}

The entropy contributions obtained from the three CFs, along with the OP, are shown in Fig.~\ref{fig:hex vs rho} for varying densities at constant noise amplitude, and in Fig.~\ref{fig:hex vs eta} for varying noise amplitudes at constant density. The OP increases  with increasing density, as more and more particles are located within the interaction radius. The OP decreases with increasing noise amplitude, as particle orientations become less and less correlated. These are the expected trends, in line with the common view of the order-disorder transition in the Vicsek model as arising from a competition between mutual alignment and noise.
To establish the existence of a phase transition and characterize it (\eg whether it is first- or second-order) requires more than these monotone trends in the OP. One should demonstrate numerically the existence, in the thermodynamic limit, of a critical noise $\eta_c (\rho)$ and critical density $\rho_c (\eta)$ such that for $\eta > \eta_c$ or $\rho < \rho_c$ the OP vanishes. The order of the transition can be inferred from the presence or absence of a jump in the OP, the extraction of critical exponents~\cite{vicsek1995model}, or Binder cumulants~\cite{gregoire2004onset}. Indeed, there is still uncertainty concerning the order of the transition in the Vicsek model depending on the implementation details (\eg intrinsic vs.\ extrinsic noise)~\cite{vicsek1995model,gregoire2004onset,chate2008modeling,vicsek2012collective}.

The excess entropy contributions give a richer account of the transition. They provide an alternative point of view concerning the interplay between positional and orientational DOFs. Consider the dependence of entropy on density at fixed noise amplitude, Fig.~\ref{fig:hex vs rho}. Alignment reduces the entropy of the orientational DOF. At low densities, the alignment requires clustering, which reduces also the positional entropy. Thus stronger alignment and tighter clusters at low density make the entropy decrease with density. At high density, on the other hand, alignment no longer requires clustering, as every particle has many neighbors even in an ideal-gas distribution. The system can ``afford" a broad distribution of particle positions without affecting much the alignment, and the entropy increases with density. (Recall that Figs.~\ref{fig:hex vs rho} and~\ref{fig:hex vs eta} show the {\em excess} entropy over that of the ideal gas, which removes the trivial reduction in entropy due to the decreasing system size $L$.)
Put together, these two trends lead to a well-defined critical density where the slope of $h_{\rm ex}(\rho)$  changes sign. As seen in Fig.~\ref{fig:hex vs rho}, this point marks the transition much more sharply than the OP, especially for small system sizes. The minimum in $h_{\rm ex}(\rho)$ is clearly observed even for $N=100$ (not shown). These transitions involve the formation of traveling bands~\cite{vicsek07model,ginelli2016}, appearing much more prominently for $N=10^4$ (not shown). In our simulations, the bands moved strictly perpendicular to the box boundaries, suggesting that they may have been boundary-related. Nevertheless, we find that the entropy is lower still for the small-but-dense clusters than for the traveling-bands phases; hence the steady rise of entropy with increasing $\eta$ in Fig.~\ref{fig:hex vs eta}. The continuous entropy minimum as a function of density, and its sharpening with increasing $N$ (not shown), support a second-order transition, at least within the range of parameters studied here.

All entropy contributions shown in Figs.~\ref{fig:hex vs rho} and~\ref{fig:hex vs eta} increase with noise amplitude, as expected. Yet, the different contributions do not increase to the same extent. Since the noise acts directly on the orientational DOF, the distribution of this DOF is affected by changes in noise amplitude more strongly than that of the translational DOF. Consequently, at a certain noise amplitude, the entropy contributions from the positional and orientational DOFs switch order. Compare  the panels for $\eta=0.4$ and $\eta=0.6$ in Fig.~\ref{fig:hex vs rho}. This crossover from an orientation-dominated regime to a translation-dominated regime does not appear to coincide with the order-disorder transition itself (see Fig.~\ref{fig:hex vs eta}), implying the existence of another transition. This may be related to the previously reported two transitions, one being the well-known flocking transition (here consistent with a second-order transition~\cite{vicsek1995model}), and the other indicating the gradual loss of orientational order while keeping some positional order (i.e., transition from homogeneous patches to bands). See, \eg the two curves in Fig.~2(e) of Ref.~\cite{ginelli2016}, and the two critical noises $\eta_{b,c}$ in Fig.~7 of Ref.~\cite{cavagna2021ent}. The latter work inferred the transition from entropy through a time-interlaced estimation algorithm, \ie including temporal information.  For the range of parameters and system sizes studied here, the crossover occurs in the range $\eta=0.4$--$0.5$, independent of density. 

Figures~\ref{fig:hex vs rho} and~\ref{fig:hex vs eta} show also the entropy contribution derived from the mixed CF, $\vecM(\vecq)$. This contribution is always more negative than the sum of the ones obtained from $S(\vecq)$ and $\vecD^1(\vecq)$ separately, as mentioned in Sec.~\ref{sec:methods}. In the Vicsek model we find that the excess entropy from $h_\mathrm{ex}[\vecM]$ is only about $1$ percent lower than $h_\mathrm{ex}[S]+h_\mathrm{ex}[\vecD^1]$. Thus, in this model, cross-correlations between particle positions and orientations do not contribute appreciably to the entropy. 

\section{Swarming bacteria}
\label{sec:bacteria}

\begin{figure}
\centering
\includegraphics[scale=1.3,trim={0 20.5cm 10cm 0},clip]{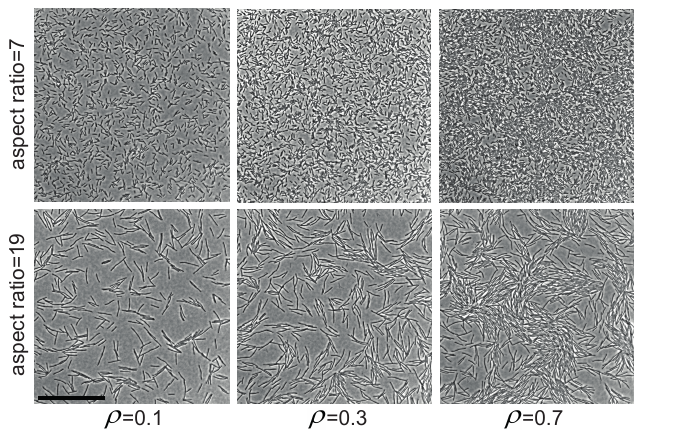}
\caption{\label{fig:bact_exp}Experimental snapshots showing swarming {\it Bacillus subtilis} bacteria with different aspect ratios and at different area fractions. Reproduced from Ref.~\cite{GilBacteria20}. The scale bar is 50$\upmu$m.}
\end{figure}

Bacterial swarming is a rapid mass-migration in which thousands of cells move collectively, forming coherent patterns. Physically, swarming is a natural example of active particles consuming energy to generate collective motion. Accordingly, understanding the constraints that physics imposes on these dynamics is essential for clarifying the mechanisms underlying swarming \cite{jeckel2019learning,be2019statistical}. Here, we revisit experiments on swarming {\it Bacillus subtilis} cells with different aspect ratios and at different area fractions\,---\,two physical parameters known to affect collective behavior~\cite{GilBacteria20}. See Fig.~\ref{fig:bact_exp}. 

Previous analysis of these experiments revealed three qualitatively distinct swarming regimes, describing how cell shape and population density govern the dynamical characteristics of the swarm. 
Strains with small aspect ratios, such as the wild-type (WT), exhibited rapid mixing,   homogeneous density, and no preferred direction of motion, for all tested values of area fraction. The absence of qualitative differences over the range of area fractions suggests a single phase. Long mutants showed different swarming statistics. In particular, two phases were identified\,---\,a dilute phase consisting of small moving clusters, and a denser phase, in which large moving clusters of the size of the viewing area caused strong number fluctuations in time \cite{GilBacteria20}. 
Importantly, the differences between the phases were identified using the cluster size distribution\,---\,an {\it ad hoc} measurement which is not well defined (the definition of a bacterial cluster is not unique)\,---\,and the temporal dynamics. Spatial correlation functions were not indicative of a phase transition. 

In Ref.~\cite{GilBacteria20}, a custom algorithm enabled tracking of individual cell trajectories and orientations. We use the same data to obtain the CFs: the structure factor $S(\vecq)$, orientations CF $\vecD^2(\vecq)$ (correlations between the nematic DOFs of the rod-like bacteria\footnote{Rod-like Bacteria such as {\it Bacillus subtilis} swarm cells are usually described as polar particles, rather than nematic ones \cite{zhang2010GNF,jeckel2019learning,be2019statistical}. However, analyzing single images, one cannot distinguish the bacterial `head' from its `tail'. For this reason, we treat cells as nematic particles.}), and their mixed CF $\vecM(\vecq)$. The entropy contributions of the corresponding DOFs are then inferred according to Sec.~\ref{sec:methods}. We concentrate on the WT and the longest mutant, with aspect ratios of 7 and 19, respectively. 

\begin{figure}
\centering
\includegraphics[scale=0.82]{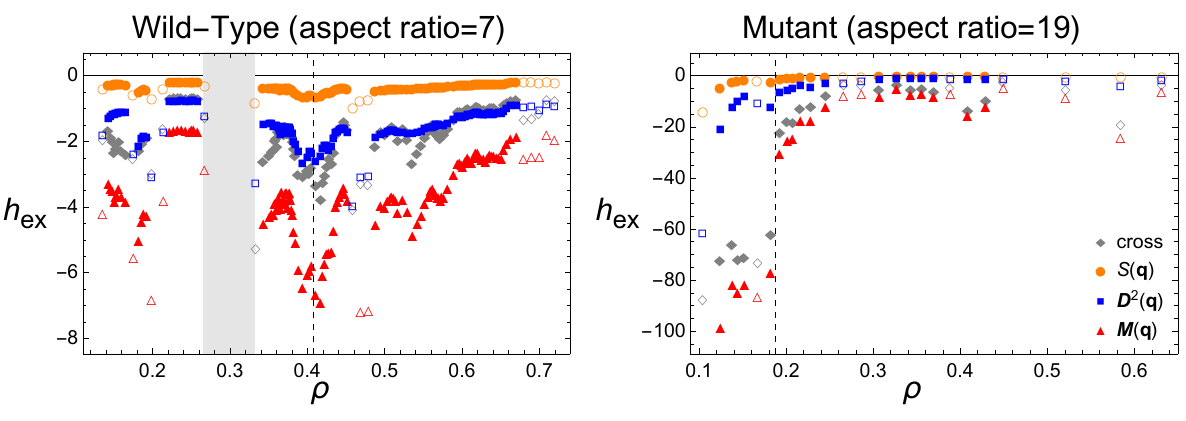}

\caption{\label{fig:bact_htot}Excess entropy per particle $h_\mathrm{ex}$ as a function of area fraction $\rho$ for the wild-type (left) and mutant (right) bacteria. Different symbols show $h_\mathrm{ex}$ as obtained from the different correlation functions: structure factor $S(\vecq)$ (orange circles), nematic orientations correlation function $\vecD^2(\vecq)$ (blue squares), and mixed correlation function $\vecM(\vecq)$ (red triangles). Also shown is the net entropy of position-orientation cross-correlations (gray diamonds). Full symbols correspond to points obtained from 250--300 samples, and empty symbols to less reliable points obtained from less than 250 samples.  The dashed vertical line in the right panel marks a first-order transition. The gray area and dashed vertical line in the left panel may indicate hitherto unrecognized transitions.}
\end{figure}

Figure~\ref{fig:bact_htot} shows the excess entropy per particle, $h_\mathrm{ex}$, as inferred from the three CFs mentioned above, as a function of the area fraction, for the WT (left panel) and long mutant (right panel). We plot also the entropy contribution of cross-correlations, $h_\mathrm{cross}=h_\mathrm{ex}[\vecM]-(h_\mathrm{ex}[S]+h_\mathrm{ex}[\vecD^2])$.

For the elongated  mutant (Fig.~\ref{fig:bact_htot}, right), a sharp jump in all entropy contributions around $\rho=0.18$ indicates a first-order transition. (The relative jump is similar for all four entropies; it is less evident in Fig.~\ref{fig:bact_htot} for the smaller contributions from $S(\vecq)$ and $\vecD^2(\vecq)$.) This agrees with the observations in  Ref.~\cite{GilBacteria20}. The transition from a lower excess entropy in the dilute phase to a higher one in the dense phase is counter-intuitive. One expects  a higher area fraction to lead to stronger alignment and thus to lower entropy. As we have seen in the Vicsek model, however, this is not always the case. At low area fractions isolated cells are immobile, possibly because single cells cannot form a hydrated layer necessary to move \cite{be2019statistical}. As a result, the alignment of cells that are close to one another (and therefore moving together) is significantly higher,  resulting in strong position-orientation cross-correlations. Indeed, these cross-correlations give the dominant contribution to the entropy (gray diamonds in Fig.~\ref{fig:bact_htot}, right panel). The first-order transition implies coexistence of the two states\,---\, stationary isolated cells and small moving clusters. The motion of the clusters makes the two `phases' mix spatially.

For the WT (Fig.~\ref{fig:bact_htot}, left), the entropy reveals subtle phenomena and possibly new transitions, which the standard CF analysis did not  identify~\cite{GilBacteria20}. First, there is a gap between $\rho=0.27$ and $\rho=0.33$, reflecting a very small number of samples in this range, which did not allow entropy calculation.\footnote{The data are sufficient for inferring a correlation length \cite{GilBacteria20}.} The entropy contributions of alignment and cross-correlations show a jump across the gap, possibly indicating a first-order transition. These observations may again be rationalized by the coexistence of two phases. Unlike the elongated mutant, WT cells can move at low area fraction, and their spatial distribution is locally homogeneous at all area fractions. Hence, the swarm segregates into locally homogeneous, mobile domains of low and high area fraction. Samples whose mean area fraction falls in the coexistence range would be found only when the field of view covers an intermediate region separating the two phases, which explains the scarcity of such samples.

Thus the first-order transition is seen for both the WT and mutant strains; yet, because WT cells can move at low area fraction while the mutants cannot, the transition is manifested differently. The mutant undergoes the transition  
at a lower area fraction. This may hint at an active isotropic-to-nematic transition, which is promoted by the mutant's large aspect ratio.

The second interesting observation concerning the WT cells is the sharp minimum at $\rho=0.41$, which may indicate a second-order transition. The non-monotone dependence on area fraction is manifested in all entropy contributions. It is reminiscent of the entropy variation across the flocking transition in the Vicsek model (cf.\ Fig.~\ref{fig:hex vs rho}), involving a competition between positional and orientational entropies.

\section{Discussion}
\label{sec:discuss}

This work has two goals. The physical goal has been to identify and characterize hitherto unrecognized features of two canonical active systems\,---\,the Vicsek model and bacterial swarming. A more general, technical goal has been to demonstrate the applicability and advantages of the entropy-bounds method. 

In the Vicsek model we have found further evidence for an order-disorder transition of continuous (second-order) nature. We have clarified the interplay between positions and orientations before and after the transition while pin-pointing its critical density for varying noise amplitude. Our method has also allowed the identification of a crossover from orientation-dominated to position-dominated collective dynamics above a certain noise amplitude (consistent with previous studies~\cite{ginelli2016,cavagna2021ent}), which is insensitive to density. 
In bacterial suspensions we have found clear-cut evidence for a discontinuous, first-order transition, associated with separation into coexisting domains of different density and mobility. At higher area fraction we have identified another transition, continuous (probably second-order) in nature, with entropy dependence resembling the Vicsek flocking transition.

Concerning the second goal, we have established the ability of the entropy-bounds method to distill the information contained in structural correlations into a single useful number while resolving the different contributions coming from different microscopic DOFs. This ability has led to the new observations mentioned above. 
The method has another crucial advantage which has not been used here. We have measured the CFs from `microscopic' configurations obtained by simulation or microscopy. These configurations could be fed into alternative, computational entropy-estimation methods. However, CFs are coarse-grained properties which can be obtained by macroscopic measurements without direct access to the particles' DOFs. For example, the CFs could be obtained by scattering and birefringence. Thus we hope that the entropy-bounds method will become a wide-spread useful tool for identifying and characterizing transitions of diverse systems in and out of equilibrium.

\begin{acknowledgments}
  The research has been supported by the Israel Science Foundation (Grant no.\ 986/18).
\end{acknowledgments}

\end{document}